\documentclass[a4paper,11pt]{article}
\pdfoutput=1 

\usepackage{jcappub} 

\usepackage[T1]{fontenc} 
\usepackage{multirow}
\usepackage{xcolor}

\title{\boldmath Improving cosmological covariance matrices with machine learning}

\author{Natalí S. M. de Santi}
\author{and L. Raul Abramo}

\affiliation{Instituto de Física, Universidade de São Paulo,\\ R. do Matão 1371, 05508-900, São Paulo, Brazil}

\emailAdd{natalidesanti@gmail.com}

\abstract{
  Cosmological covariance matrices are fundamental for parameter inference, since they are responsible for propagating uncertainties from the data down to the model parameters. However, when data vectors are large, in order to estimate accurate and precise covariance matrices we need huge numbers of observations, or rather costly simulations - neither of which may be viable. In this work we propose a machine learning approach to alleviate this problem in the context of the covariance matrices used in the study of large-scale structure. With only a small amount of data (matrices built with samples of 50-200 halo power spectra) we are able to provide significantly improved covariance matrices, which are almost indistinguishable from the ones built from much larger samples (thousands of spectra). In order to perform this task we trained convolutional neural networks to denoise the covariance matrices, using in the training process a data set made up entirely of spectra extracted from simple, inexpensive halo simulations (mocks). We then show that the method not only removes the noise in the covariance matrices of the cheap simulation, but it is also able to successfully denoise the covariance matrices of halo power spectra from N-body simulations. We compare the denoised matrices with the noisy sample covariance matrices using several metrics, and in all of them the denoised matrices score significantly better, without any signs of spurious artifacts. With the help of the Wishart distribution we show that the end product of the denoiser can be compared with an effective sample augmentation in the input matrices. Finally, we show that, by using the denoised covariance matrices, the cosmological parameters can be recovered with nearly the same accuracy as when using covariance matrices built with a sample of 30,000 spectra in the case of the cheap simulations, and with 15,000 spectra in the case of the N-body simulations. Of particular interest is the bias in the Hubble parameter $H_0$, which was significantly reduced after applying the denoiser.\\

  \hspace{-0.9cm}{\bf Keywords:} cosmological simulations, power spectrum, cosmological parameters from LSS, machine learning.  

}

\begin{document}
\maketitle
\flushbottom

\section{Introduction}
\label{sec:intro}

The end product of cosmological data analysis is parameter inference, which constrains our theoretical models.
Covariance matrices lie at the connection between theory and data: they quantify the amount by which we expect our measurements to fluctuate, given the underlying physical phenomena as well as the conditions under which we collect those data sets. 

In the context of galaxy surveys, the physical phenomena include the Gaussian random nature of the initial density fluctuations, and the observational conditions include the total volume of the survey, the shape of that volume (i.e., the mask), the mean density of galaxies that were detected, and possibly many other real-life factors.
Although analytical or semi-analytical approaches can sometimes provide a good first approximation for the covariance matrix, both the physical models and the observational conditions are often best represented in terms of simulations.
Given a sample of independent simulations and some set of summary statistics that we would like to analyze, we then compute the sample covariance, which should reproduce the 
statistical errors of those statistics, leading to constraints on the underlying physical models.
However, sample covariance matrices must be both precise and accurate \cite{hartlap2007}, which means that not only our models for the physical phenomena and data acquisition need to fulfill some requirements, but that we need large samples as well, otherwise we run the risk of biasing our parameter estimation \cite{dodelson2013, taylor2013}.

The estimation of increasingly large and complex covariance matrices is therefore a key challenge for experiments which detect huge numbers of tracers distributed over vast cosmological volumes -- as is the case, e.g., for the Dark Energy Survey (DES) \cite{DES2005}, Javalambre PAU Astrophysical Survey (J-PAS) \cite{JPAS2014}, Prime Focus Spectrograph (PFS) \cite{PFS2014}, Dark Energy Spectroscopic Instrument (DESI) \cite{DESI2016}, Large Synoptic Survey Telescope (LSST) \cite{LSST2008} and Euclid \cite{Euclid2013, Euclid2011}. 
Trying to reproduce the observables in those surveys implies a series of assumptions: besides the cosmological model and the properties of the tracer populations with respect to the dark matter halos (the halo-galaxy connection for the tracers in that survey), one must also know precisely the mask, the selection function of the tracers, the depth of the images or spectra, as well as the probability distributions for object classification and photometric redshifts, if applicable.
These systematic effects are usually addressed via simulations that mimic the properties of the catalogues that are derived from the survey.

However, the number of simulations necessary to properly characterize these effects, and to allow for unbiased estimation of the parameters, 
is often very large, which represents a daunting computational challenge \cite{heavens2017},
especially when it is important to properly model the non-linear scales \cite{Blot2016}. 
Using the power spectrum as our summary statistics, the sample size (the number of simulations) that is
typically required to fulfill those needs is around $N_s \sim n_k^2$, where $n_k$ is the number of $k$
bins (bandpowers), and this number can grow even more with different tracers of the large-scale
structure and all the resulting auto- and cross-spectra.
This number is now under pressure from two sides: on one hand, in order to test the physical phenomena of interest we need to increase the dynamic range of our surveys to both larger and smaller scales, but without losing resolution -- and that means more bandpowers. 
On the other hand, astrophysical surveys are increasingly able, either by themselves or in combination, to map the universe with much greater completeness by detecting multiple tracers of the large-scale structure.
Therefore, it is of paramount importance to optimize methods that can estimate efficiently, and with greater precision, these cosmological covariance matrices.

Several efforts have already been made with the goal of obtaining precise covariance matrices using smaller samples -- for a review see Ref. \cite{taylor2013}. 
This problem does not exist, of course, if we employ analytical approximations that try to codify the impact on non-linear clustering \cite{Meiksin1999}.
Analytical models can also help to enhance covariance matrices, e.g., by using $\chi^2$ distributions from simulations, leading to a reduction in the number of realizations needed to achieve a certain threshold in accuracy \cite{alessandra2022}.
In the absence of a large sample, a commonly used technique is the Jackknife method \cite{Tukey1958, Efron1980}, which relies on sub-sampling an original data set. 
Another possible direction is data compression, which actually means reducing the size of the data vector by maximizing the Fisher information, which can be done both in the context of parameter-independent covariances \cite{Heavens2000}, as well as parameter-dependent ones \cite{heavens2017}. 
A closely related method relies on reducing the dimensionality of the observables (the parameters), which then allows for a less noisy estimation of covariance matrices given the same sample size \cite{Philcox2021}.
It is also possible to resample some specific modes or parts of the data vector \cite{Schneider2011}.

Last but not least, we have approximate numerical methods, which in the context of large-scale structure are provided by PTHalos \cite{Scoccimarro2002}, EZMocks \cite{Chuang2015}, PINOCCHIO \cite{PINOCCHIO2013}, PATCHY \cite{{Kitaura2014}}, HALOGEN \cite{HALOGEN2015}, Lognormal \cite{Lognormal2017}, ICE-COLA \cite{Izard2016}, ExSHalos \cite{ExSHalos2019}, BAM \cite{BAM2019, BAM2020} and many other techniques, which all attempt at generating halo catalogues using semi-analytical approximations or by emulating much more expensive N-body simulations.
However, even if the results seem compatible with the numerical simulations, as shown in Refs. \cite{Lippich2019, Blot2019, Colavincenzo2019}, covariance matrices derived from these approximation schemes can lead to statistical deviations of  up to $\sim$ 5\% for the bias in the estimation of cosmological and
nuisance parameters, and of around $\sim$ 10 \% for the volumes in parameter space, when compared with
the true ($N$-body) estimations, which may fall short of the accuracy required for precision cosmology
from future surveys.
Nevertheless, it is possible to use mocks in order to reduce the number of simulations needed for characterizing the statistics of the matter power spectrum or bispectrum, by exploiting the correlations between the mocks and N-body simulations -- see, e.g. CARPoll \cite{Chartier2021, Chartier2022}.

From a different perspective, machine learning (ML) techniques offer alternative solutions to some of these challenges.
ML methods are able to quickly predict non-linear relations in complex (and large) data sets, from tabular, images or even N-dimensional data. 
There are efforts trying to speed-up the process of producing high-resolution N-body simulations, by starting from lower-resolution ones, and letting the ML fill in the detailed structures on small scales \cite{Li2021, Ni2021, Ramanah2020}. 
In Refs. \cite{He2019, renan2020, Kaushal2021} the authors try to emulate the full non-linear evolution of N-body simulations by inputting only approximate simulations of these.

Another approach employs ML to perform parameter-free inference, by taking data directly from the simulations (without the need of a summary statistics and, thus, model comparison). 
Recent papers have shown competitive results compared with traditional statistical inference methods. 
In Ref. \cite{Ravanbakhsh2017} the authors used 3D convolutional networks in a volumetric representation of dark matter simulation to directly derive the cosmological parameters. 
On another level, using hydrodynamic simulations from CAMELS, the authors of Ref. \cite{Villaescusa-Navarro2021} showed how to infer not only cosmological but also astrophysical parameters using 2D maps of multiple fields such as projections of the gas temperature and of the dark matter density. 
In particular, in Ref. \cite{Villaescusa-Navarro2022} it was shown how neural networks can be used to derive a number of parameters from the physical properties of a single galaxy.

In this paper we propose a new approach, that employs convolutional neural networks (CNN), more specifically image denoising techniques, as a tool to enhance sample cosmological covariance matrices that are based on a small number of high-resolution simulations. 
As we will show, the final covariance matrices (after denoising) become as precise and accurate as the ones obtained with a much higher number of high-resolution simulations.
The idea is to train the ML method using data coming from halo mock generators, and then to apply that machinery to improve (``denoise'') a sample covariance matrix that was constructed using a small sample of very accurate, high-fidelity N-body simulations.
In practice, we train the ML denoiser using covariance matrices for power spectra from halo mocks produced by ExSHalos \cite{ExSHalos2019}, and then we apply the denoiser to covariance matrices produced using matching halo catalogues extracted from the Quijote suite of N-body simulations \cite{QUIJOTE2020}.
This process shows, first, the power of generalization of the method, which is able to improve covariance matrices from a set of simulations that the ML has never seen before.
And second, that the ML denoiser can generate  cosmological covariance matrices that are in all respects equivalent to those produced from thousands of N-body simulations.

This paper is organized as follows: in section \ref{sec:cats} we describe the halo catalogues, starting with the mocks used to build the ML model, the corresponding N-body catalogues, and the procedure that we used to match the two.
Section \ref{sec:met} comprehends the methodology, where we specify the putative survey we considered and the chosen summary statistics (the power spectrum),
the nuisance parameters (in our case, the bias),
the computation of the covariance matrices themselves, 
as well as other details about the ML technique that we have used.
In section \ref{sec:results} we present the main results of this work, focusing in the comparison of the parameter estimation that results from using the different covariance matrices in Markov Chain Monte Carlo (MCMC) explorations of the likelihood.
We also show the matrices themselves for visualization purposes, and compare them using metrics such as the mean squared error (MSE).
In order to provide a quantitative idea about the level of improvement that results from applying the ML denoiser we also compare the properties of the matrices with those obtained using the Wishart distribution, to show that the denoiser acts in a way that is analogous with an augmentation of the sample size of the covariance matrices.
Finally, in section \ref{sec:conc}, we discuss the implications of our results and explain future applications.

\section{The halo catalogues}
\label{sec:cats}

In this paper we have used two different halo catalogues: the ones obtained from a halo mock generator, called ExSHalos, are the ``cheap'' simulations, and the ones from the Quijote suite are the ``high-fidelity'' N-body simulations. In this section we briefly explain some of their main features, and our method for matching them so there is greater compatibility between the data sets.

\subsection{ExSHalos}

ExSHalos (Excursion Set Halos) constitutes a new, simple, fast, and parameter-free method to generate dark matter halo catalogues \cite{ExSHalos2019}. 
ExSHalos basically implements the notion of excursion sets \cite{Bond1991, Maggiore2010}, and then it corrects the positions of the peaks using Lagrangian perturbation theory (LPT) \cite{Vlah2015, Matsubara2008}.
The method requires as inputs only a fiducial cosmology, the linear matter power spectrum, and the threshold density for halo formation in linear theory (constant or ellipsoidal collapse barriers). 
For the ExSHalos mocks we used a linear power spectrum from the Code for Anisotropies in the Microwave Background (\href{https://camb.info/}{\texttt{CAMB}}) \cite{CAMB2011}. 
We have chosen the constant barrier and used LPT to second order. The cosmology was chosen according to the standard one in the Quijote suite, as well as the size of the box
[of volume $\left( 1000 \, {\rm Mpc}/h\right)^3$],
for which we used cubic cells of 1 $({\rm Mpc/h})^3$ volume at a fixed redshift $z = 0$. In total, we have produced 30,000 mock catalogues (hereafter, $N_{max} = 30,000$ for ExSHalos), but it should be noticed that not all the ML models used in this work needed to use this amount.

\subsection{Quijote}

The Quijote suite is a set of 43,100 full N-body simulations designed to provide a very large data set of cosmological simulations for ML applications, and also aiming at the quantification of the information content about cosmological observables \cite{QUIJOTE2020}. 
The simulations employed dark matter-only particles using the TreePM code GADGET-III -- the third generation of the well known GADGET-II algorithm \cite{GADGET-2-2005}.
Initial conditions were generated at a redshift $z = 127$,
using an input matter power spectrum and a matter transfer function computed with the help of \texttt{CAMB}, and were evolved up to $z = 0$. 
All the Quijote simulations have a volume of $(1000 \, {\rm Mpc}/h)^3$, with 512$^3$ cold dark matter particles. 
The suite has simulations for a range of cosmological models, but the main (standard) fiducial cosmology follows the Planck best-fit model \cite{planck2018}: $\Omega_m = 0.3175$, $\Omega_b = 0.049$, $h = 0.6711$, $n_s = 0.9624$, $\sigma_8 = 0.834$, $M_{\nu} = 0.0 eV$ and $\omega = - 1$. 

We have downloaded 15,000 halo catalogues, which is the maximum number of Quijote simulations for the main fiducial cosmology (hence, $N_{max} = 15,000$ for Quijote), all at $z = 0$, with halos identified using the Friend-of-Friends (FoF) algorithm \cite{Davis1985} with linking length parameter $b = 0.2$.
Those catalogues were downloaded using the 
\href{https://docs.globus.org/cli/}{\texttt{globus}} command line interface -- however, we stress that this larger sample was only used to test and validate our ML method: all the training was performed using the ExSHalos data set.

\subsection{The match between the catalogues}
\label{sec:match}

\begin{figure}[h!]
  \centering
  \includegraphics[scale = 0.75]{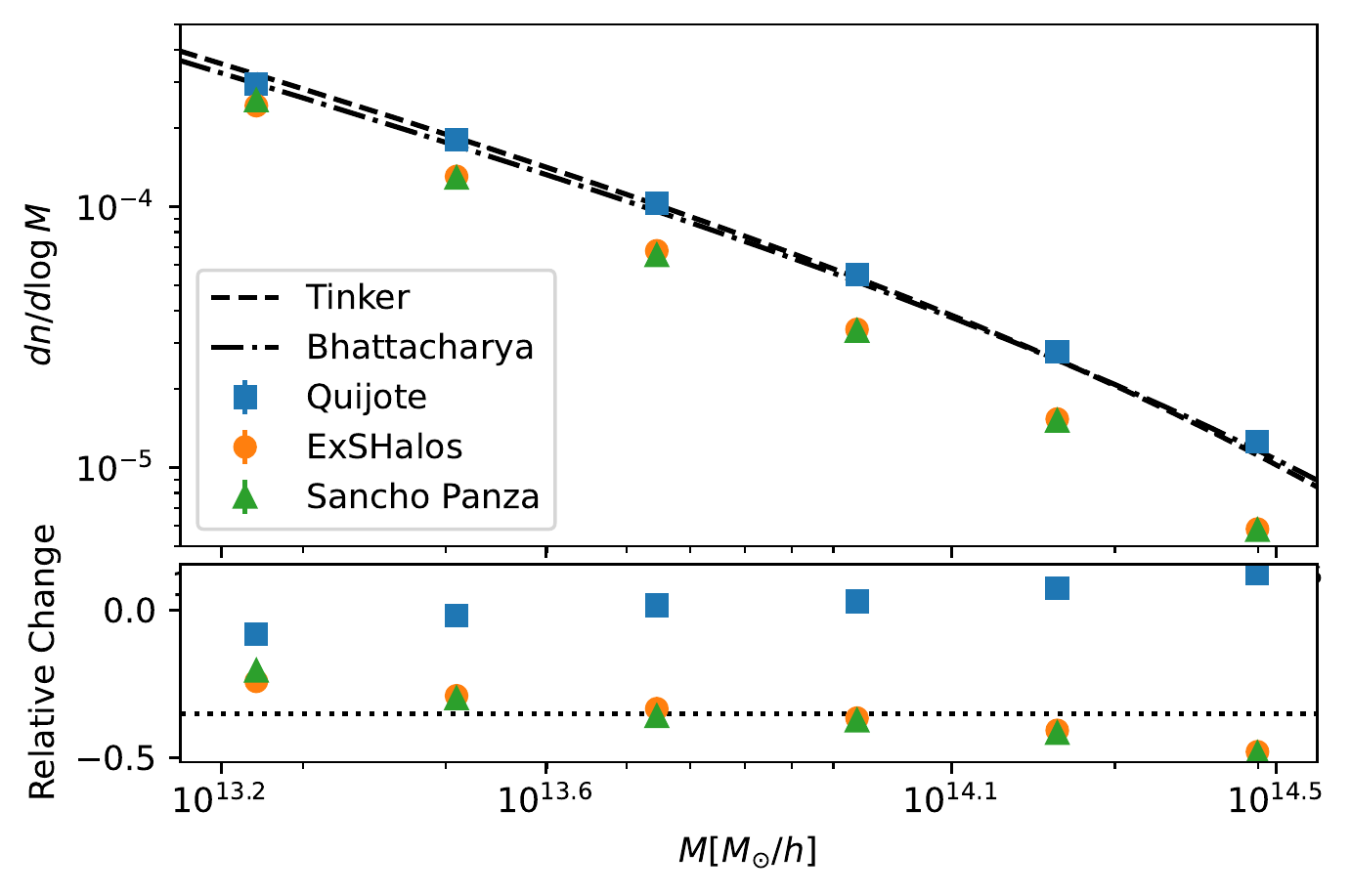}
  \caption{Halo mass functions: Quijote, ExSHalos and Sancho Panza simulations, compared to the phenomenological halo mass functions of Tinker et al. \cite{Tinker2008} and Bhattacharya et al. \cite{Bhattacharya2011}. The lower panel shows the relative change of the simulations with respect to the Tinker halo mass function.}
  \label{fig:HM_comp}
\end{figure}

In order to match the two catalogues we ensured that both Quijote and ExSHalos provided a similar numbers of halos, with approximately the same values for the halo bias.
We have analyzed the halo mass functions of both catalogues considering halos in the mass interval $M \in [10^{13.12}, 10^{14.6}] M_{\odot}/h$, in intervals of $\Delta \log_{10} M \simeq 0.25 \log_{10} (M_{\odot}/h)$, which corresponds to six bins of halo mass. 
We then compared the resulting halo mass functions from the catalogues with the fits by Tinker et al. \cite{Tinker2008} (using $\Delta = 200$ to compare with Quijote halos, found with link parameter $b = 0.2$) and Bhattacharya et al. \cite{Bhattacharya2011}, which were computed with the help of the
\href{https://bdiemer.bitbucket.io/colossus/lss_mass_function.html}{Colossus} library \cite{Colossus}.

The results of this comparison are shown in figure \ref{fig:HM_comp}. 
From that figure it is clear that ExSHalos do not show a perfect agreement with the phenomenological fitting functions: the lower panel shows a relative difference with respect to Tinker's fitting function that hovers around a deficit of $\sim 35\%$ (which is represented by the dotted black line in the lower plot). 
When compared to halos from real N-body simulations (Quijote), the results agree with both the Tinker and Bhattacharya halo mass functions on almost all scales, deviating only $\sim 10\%$ at the lower mass end. 
Even if the halos in ExSHalos and Quijote had the same bias, the different abundances would have an impact on shot noise and in the covariance of the power spectra.
Hence, in order to improve the match between ExSHalos and Quijote, we randomly removed halos from each mass bin in the Quijote catalogues in such a way that their final abundances match the same halo bins in ExSHalos -- in that respect, see also Ref. \cite{Blot2019}. 
We refer to the resulting ``clipped'' Quijote catalogues as the {\em Sancho Panza} sample.

Not only the halo mass was matched, but we have tested the halo bias, for these same mass bins, to compare the results between ExSHalos and Quijote/Sancho Panza maps and we have observed that they were similar for each mass bins. The detailed comparison is presented in section \ref{sec:bias}.

\section{Methodology}
\label{sec:met}

In this section we describe how we simulate a survey with a non-trivial mask, and show the effects of that mask on the power spectrum of the tracers in the two simulations. 
We also make a preliminary check that the bias in the two samples is nearly the same, as well as the dynamical range of scales that we are able to analyze. 
At the end of the section we describe the construction of the samples and the computation of the covariance matrices.

\subsection{The power spectrum}
\label{sec:PS}

For the sake of simplicity, in this work we chose as tracers the halos belonging to the first mass bin discussed in  section \ref{sec:match}, i.e., halos with masses between $M \in [10^{13.12}, 10^{13.37}] M_{\odot}/h$, with a mean mass $\langle M \rangle = 10^{13.245} M_{\odot}/h$, both in the ExSHalos and in the Sancho Panza samples\footnote{Notice that the results presented here do not depend significantly on our choice of tracer, especially for the parameter estimation. 
Nevertheless, different tracers have different cosmological covariance matrices: e.g., shot-noise may affect the diagonal and off-diagonal terms of the covariance matrices in different ways.}.
This choice minimizes the differences in the halo mass functions of the two samples, which are also closer to the mass functions by Tinker and Bhattacharya for that mass bin.

\begin{figure}[h!]
  \centering
  \includegraphics[scale=0.75]{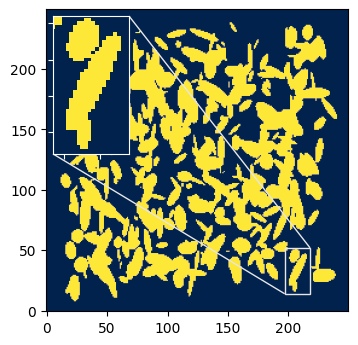}
  \caption{Slice of the mask of random ellipsoids. The inset shows a zoom in on three ellipsoids, for a better visualization.
  \label{fig:mask}}
\end{figure}

The power spectrum is our chosen summary statistics, so that for each simulation the data vector corresponds to a set of $P(k)$, and that is what we use to build the covariance matrices. 
In order to simulate real-life effects and power spectra which are closer to the ones measured in realistic cosmological surveys, we have masked the halo maps in a way that attempts at emulating a mask covering some regions of the sky: this mask reflects the survey's footprint as well as bright stars, cloudy nights, regions with poor seeing, etc \cite{mask2018}. 
The main point of using a mask, in the context of this work, is to induce non-trivial correlations between different spectral modes, in a way that resembles real surveys.

Our mask was built using a single realization of randomly placed ellipsoids with random sizes and
orientations, in such a way that regions outside those ellipsoids were masked out (i.e., regions outside the mask are assigned weight zero, while regions inside it have weight 1).
The mask occupies approximately $\sim 50\%$ of the total box volume, and we estimated the spectrum on a grid of the entire box, with cells of $4 \, h^{-1} \, {\rm Mpc}$ on a side.
Figure \ref{fig:mask} shows a slice of our mask, with the sizes and shapes of the ellipsoids as well as shapes of the mask edges with respect to the cell size that we used in our computations.

\begin{figure}[h!]
  \centering
  \includegraphics[scale = 0.75]{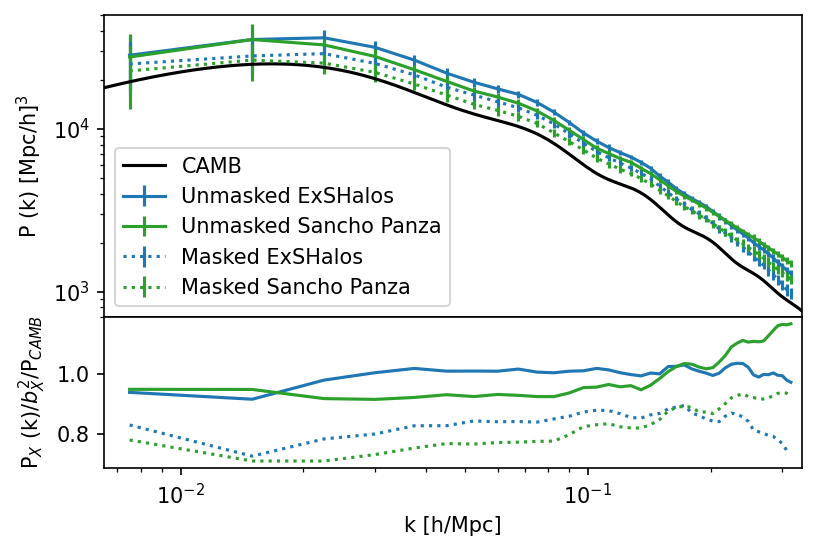}
  \caption{Power spectra of the halos from ExSHalos (blue lines) and Sancho Panza (green), masked (dotted) and unmasked (solid). 
  The linear matter power spectrum is also shown as the solid black line for comparison.}
  \label{fig:PS_example}
\end{figure}

We computed the spectra of both the masked and unmasked catalogues, which are shown (in terms of the mean for 100 maps) in figure \ref{fig:PS_example}. 
It can be seen that the shapes of the power spectra of the ExSHalos and Sancho Panza samples are similar on scales larger than $k \lesssim 0.15 \, h \, {\rm Mpc}^{-1}$. 
On smaller scales, although the shot noise for both samples are identical, the differences start to become more evident, and are due to the approximate treatment of structure formation in halo mocks such as ExSHalos.
The effect of the mask can also be seen from this plot, in terms of a suppression of the clustering amplitude on all but the smallest scales.

\subsection{The bias}
\label{sec:bias}

In order to check the similarity between the ExSHalos and Sancho Panza halo catalogues we also computed the bias of the halos in our chosen mass bin. We have used the bias definition  \cite{Tinker2010}:
\begin{equation}
  b (k) = \sqrt{ \frac{ P_{h} (k) }{ P_{lin} (k) } } \; ,
\end{equation}
where $P_{h} (k)$ represents the halo power spectrum and $P_{lin} (k)$ the linear matter power spectrum.
The bias was computed for 100 ExSHalos and 100 Sancho Panza catalogues, and averaged over scales in the range $k \in [0.015, 0.2625]$ $h$ Mpc$^{-1}$, where the small-scale cut-off corresponds to the value of $k$ for which shot noise becomes 80\% of the power spectrum \cite{Valcin2019}.

The results of this comparison are presented in table \ref{tab:bias}, where besides the mean and standard deviation of the halo bias for our maps we also show the expected bias obtained with the Tinker \cite{Tinker2010} and Bhattacharya \cite{Bhattacharya2011} fits.
We should stress that, when we perform our cosmological parameter inference, we treat the bias as a nuisance parameter -- see section \ref{sec:results}.

\begin{table}[h!]
 \caption{\label{tab:bias} Bias for the halo catalogues.}
 \begin{center}
  \begin{tabular}{cccc}
   \cline{1-4}
   \textbf{Catalogue} & \textbf{Bias} & \textbf{Tinker} & \textbf{Bhattacharya} \\
   \cline{1-4}
    ExSHalos & 1.245 $\pm$ 0.037 & \multirow{2}{*}{1.208} & \multirow{2}{*}{1.096} \\
    Sancho Panza & 1.223 $\pm$ 0.057 & & \\
   \cline{1-4}
   \end{tabular}
  \end{center}
\end{table}

\subsection{The data set of the covariance matrices}
\label{sec:dataset}

All the cosmological covariance matrices in this paper were built by computing the sample covariance for the power spectra in the data vector as:
\begin{equation}
  Cov^{(N)} [ P (k_i), P(k_j) ] = \frac{1}{(N - 1)} \sum^{N}_{l = 1} \left[ P (k_i)_l - \bar{P} (k_i) \right] \, \left[ P (k_j)_l - \bar{P} (k_j) \right] ,
\end{equation}
where $N$ is the number of spectra in the data vector (the sample size), $P (k_i)_l$ is the value of the $l$th spectra for the $i$th bin, and $\bar{P} (k_i)$ is the mean power spectrum.
We should regard any sample covariance as a random matrix: given two different samples of $N$ data vectors, the two sample covariances will be different by an amount that reflects the level of statistical fluctuations in the two samples.
Larger samples are naturally less subject to those fluctuations, but they require more simulations.
Our goal here is to show that ML techniques are able to identify (and correct) at least part of those patterns of statistical fluctuations, even if they are trained using simplified simulations.

The data set of sample covariance matrices used to train the ML suite was built using only the spectra from the ExSHalos catalogues, whereas the data set used for the final validation of the predictive power of the suite was built using spectra from Sancho Panza.
In order to train, validate and test our ML denoiser we used samples of different sizes. 
First, we constructed the {\em input} covariance matrices, which correspond to small sample sizes
($n$ spectra), and are the ones we would like to enhance with our denoiser.
Next, we computed covariance matrices with larger sample sizes ($N$ spectra, the {\em target} matrices), which serve to teach the ML denoiser about how to clean the noise of the input matrices by performing the task $n \to N$.
Finally, we compare the cleaned version of the input matrices with the {\em best} possible covariance matrix, which is computed using the maximum sample size ($N_{max}$ spectra), corresponding to the total entire set of catalogues.

We have used target sample covariance matrices with $N = 1000$ spectra, and input matrices with sample sizes in the interval $n \in [50, 250]$, in increments of $\Delta n = 25$.
These choices were informed both by the typical number of mocks used in the estimation of cosmological covariance matrices, and by our goal to test what is the minimum number of input spectra ($n$) that results in denoised matrices which are as accurate and precise as the ones computed with samples of $N = 1000$ spectra. 
The underlying idea being that we can start with an input covariance matrix constructed from a small number ($n$) of high-resolution N-body simulations, and denoise that covariance using a ML method trained using a high ($N$) number of simplified mocks.

We simulated a grand total of $N_{max} = 30,000$ halo maps using ExSHalos, and we downloaded a total of
$N_{max} = 15,000$ Quijote simulations, which then became (after matching with ExSHalos) our sample of
$N_{max} = 15,000$ Sancho Panza halo catalogues. 
The spectra from these very large samples were used to compute the best case scenario, corresponding to
the ideal cosmological covariance matrix for each simulation. 
It is important to say that we do not necessarily use this total number of spectra to train the ML
suites: in each training we used 120 input matrices and 120 target matrices, however only the input
matrices are completely independent. 
In other words, some targets matrices may be correlated with other targets, as a single target matrix
can include many different input matrices. 
E.g., in the cases of $n = \{ 50, 100 , 200 \}$ input spectra we used $\{ 6000, 12,000 , 24,000 \}$
spectra (of the ``cheap'' simulation, ExSHalos) for the entire training process.
Since the Quijote/Sancho Panza simulations were already available, the main limitation of our model was
the computational cost associated with running the ExSHalos simulations, but in a realistic application
of our method the cost of producing the mocks would be negligible compared with the cost of running the
N-body simulations -- and our denoiser is a tool for beating down that second, much more onerous cost.

\subsection{The ML suite}
\label{sec:ML}

The ML suite we used in this work is an image denoising algorithm. 
Mathematically, the problem of image denoising is modeled by:
\begin{equation}
  \mathbf{y} = \mathbf{x} + \pmb{\nu} \; ,
\end{equation}
where the noisy image $\mathbf{y}$ is composed of an unknown image signal $\mathbf{x}$, and the noise $\pmb{\nu}$, which is usually modeled as a Gaussian random variable with null expectation value and standard deviation $\sigma$.
The goal of the denoiser is to produce a cleaned version of the image, $\mathbf{\hat{x}}$, by reducing the noise while keeping its
original properties and features while not creating new artifacts in the process. For reviews on image denoising techniques see Refs. \cite{Milanfar2013, Fan2019, Tian2019}.

CNNs, or auto-encoders, represent the state-of-art in terms of performance for this task, mainly because they are purely data driven, with no assumption about the nature of the noise 
\cite{Mao2016, Chollet2017, Fan2019, Tian2019}, and for this reason it was the method chosen in our work. 
The CNNs take a pair of images: a noisy image $\mathbf{y}$ (input) and a clear image $\mathbf{x}$ (target).
Given many such pairs, they learn to recognize what is signal, what is noise, and how to remove that noise, predicting images which are closer (less noisy) to the ones used as target -- i.e., as close as possible to the ground truth images $\mathbf{x}$.
In the present work the noisy images are the input cosmological covariance matrices (built with samples of $n$ spectra) and the target images are the matrices built with samples of thousands of spectra.

Usually a basic auto-encoder has two parts: an encoder, followed by a decoder. 
First, the encoder takes an input image $\mathbf{y}$, of dimensions
$d \times \bar{d}$, and maps it into a hidden representation $\mathbf{z}$, of dimensions $d' \times \bar{d}$, performing a mapping
$\mathbf{z} = f_{\Theta} (\mathbf{y}) = f (\mathbf{W} \mathbf{y} + \mathbf{b})$, parameterized by $\Theta = \{ \mathbf{W}, \mathbf{b} \}$. 
$\mathbf{W}$ is a weight matrix and $\mathbf{b}$ a bias, both
with dimensions $d' \times d$.
Then, the decoder takes $\mathbf{z}$ and map it back into
$\mathbf{\hat{x}} = g_{\Theta'} (\mathbf{z}) = g (\mathbf{W}' \mathbf{z} + \mathbf{b}')$, with another weight matrix $\mathbf{W}'$ and another bias vector $\mathbf{b}'$.
Hence, the reconstructed/denoised image $\mathbf{\hat{x}}$ has the same dimensions of the input/noisy image $\mathbf{y}$.
In this way, the auto-encoder comprehends a sequence of convolutional layers that are responsible for extracting the features from the images, capturing the abstraction of their content, and then recovering the features at the end of the process. This is performed with the requirement that the following loss function is minimized:
\begin{equation}
  {\rm min}_{\Theta, \Theta'} \left[ loss \left( \mathbf{\hat{x}}, \mathbf{x} \right) \right] =
  F \left( \mathbf{y}, \Theta, \Theta' \right) \; ,
\end{equation}
where $F (\cdot)$ is the function learned by the CNN in order to remove the noise of the images, $\Theta$ and $\Theta'$ represent the set of parameters of the CNN, and $loss (\cdot)$ is the loss function
\cite{Fan2019, vincent2008}.
The idea is that the loss measures the difference between the network predictions $\mathbf{\hat{x}}$ and the target images $\mathbf{x}$, such that the minimization of the loss function optimizes the function $F$ that removes the noise.

Auto-encoders are built using 2D convolution layers (as well their transposed version), mainly
described by their number of filters (which are responsible to keep local/value information
about the image, that are translated in the depth of the output feature map of each layer) and
their kernel size (that represent the square size of the patches to be extracted from the 
input). These parameters are responsible to learn about the translation invariance of the image 
patterns, as well to give a spatial hierarchy to them. Initial random weights can be defined
for each layer, according to some function (for instance, uniform weights). The value
transformation of the layers is done by their activation function (e.g. Leaky ReLU, hyperbolic
tangent and so on). In addition to these layers, other classes of them can be used, e.g. dropout
layers, to randomly set zeros through the results between the other layers (the amount of such 
zeros are determined by their rate parameter). The training is made by back propagation of the
images through the network, which is made choosing the number of images to be propagated (batch
size) and the number of epochs to train the model and define their final set of weights 
\cite{keras2015, Chollet2017}.

Different ML models were constructed in order to deal with each combination of input and target matrices with $(n, N)$ spectra. 
The size of the data set was composed with 120 matrices (240 in total, because each input matrix had its respective target). 
For each model we monitored the loss function for the Mean Squared Error (MSE) using 40 epochs each. 
Considering the size of the sets we have used 86 matrices in the train stage, 10 matrices for validation, and 24 matrices for the test. 
The CNN used in this work has its first two 2D convolutional layers (together with the dropout layers), with 64 and 32 filters, corresponding to the encoder part, while the 2 transposed convolutional layers (respectively with 32 and 64 filters and their respective dropout layers), followed by one convolutional layer with only one filter, corresponding to the decoder.
Each convolutional layer, except for the last one, was followed by a dropout layer, for which we have chosen a rate of 0.05. 
The activation function for each internal layer was the Leaky ReLU ($\alpha = 0.001$), and for the last layers we used the hyperbolic tangent activation function. 
Finally, the batch size was 15, all the kernels have the size of 3 $\times$ 3 pixels, we have used the GlorotUniform initializer for the weights in each layer and, since we work with real-valued matrices, the input data has a single channel. 
The entire method was implemented with the help of the
\href{https://keras.io/}{keras} library \cite{keras2015}.

We did not use pooling or unpooling layers in our network, since we verified that this operation tends to discard useful image details, as was also found by Ref. \cite{Mao2016}. 
On the other hand, the use of dropout layers has indeed improved our results. Still in accordance with the findings of the aforementioned reference, the use of a small number of layers improved the model performance. 
Our choices of hyperparameters were made according to the best final performance in terms of the lowest MSE values for the training and test sets. 
We have tested other metrics (e.g. logarithm hyperbolic cosine, mean absolute error, and mean squared logarithm error), but the best results were obtained using the MSE. 
We also tested the denoising technique using a Residual Encoder-Decoder Network (REDNet) \cite{Mao2016}, but the results were significantly worse compared with our standard architecture.

Covariance matrices often reflect a hierarchy of observables with different signals and different noises.
In order to homogenize the entries of the covariance matrices in the training stage, we normalized the rows and columns of all the matrices using the diagonal of the matrix with the highest number of spectra available (i.e., $N_{max} = 30,000$ ExSHalos spectra) according to:
\begin{align}
  Cov^{(N)}_{i j} \; \to \; \frac{Cov^{(N)}_{i j}}{\sqrt{Cov^{(N_{max})}_{i i} \, Cov^{(N_{max})}_{j j}}} . \label{eq:normNbest}
\end{align}
For the ExSHalos catalogues, the normalized $Cov^{(N_{max})}_{i j}$ becomes in fact the correlation matrix for that sample.
This normalization helped achieve a faster convergence of the ML method during the training stages\footnote{We have tested different normalization methods, for instance, using directly the correlation matrices, or using normalizations from the means of the diagonal of different matrices. However, using a fixed normalization for all the matrices proved to be the best strategy for the denoising method. We have also checked that it makes very little difference which precise normalization scheme is used: the results in terms of the parameter estimation remain basically the same.
}.
We train our ML denoised on these normalized matrices, and plug back the normalization to recover the denoised covariance matrices.
At the end of the whole process we have imposed the symmetry of the covariance matrices, $Cov_{ij} \to (Cov_{ij} + Cov_{ji})/2$. 

Besides the different ML models for each combination $(n, N)$ of input and target covariance matrices, we also used four different random seeds for each model. 
Therefore, we can account at least in part for statistical variations in our results which are due to different solutions (different CNN final weights) found during the training process. 
In our tests we selected the best-performing of those solutions to apply on the input matrices from the Sancho Panza data set.

\section{Results}
\label{sec:results}

\begin{figure}[h!]
  \centering
  \includegraphics[scale=0.26]{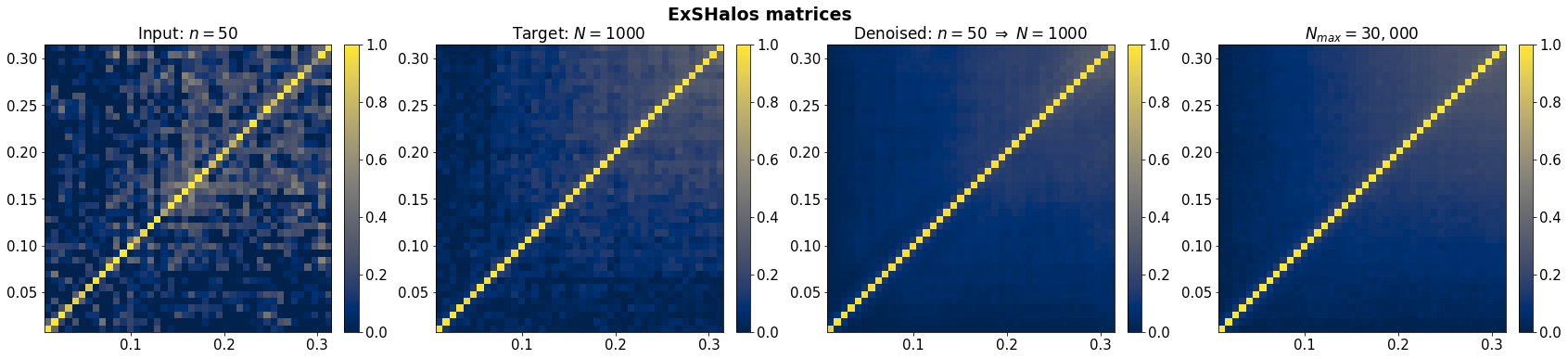}
  \vspace{0.5cm}
  \includegraphics[scale=0.26]{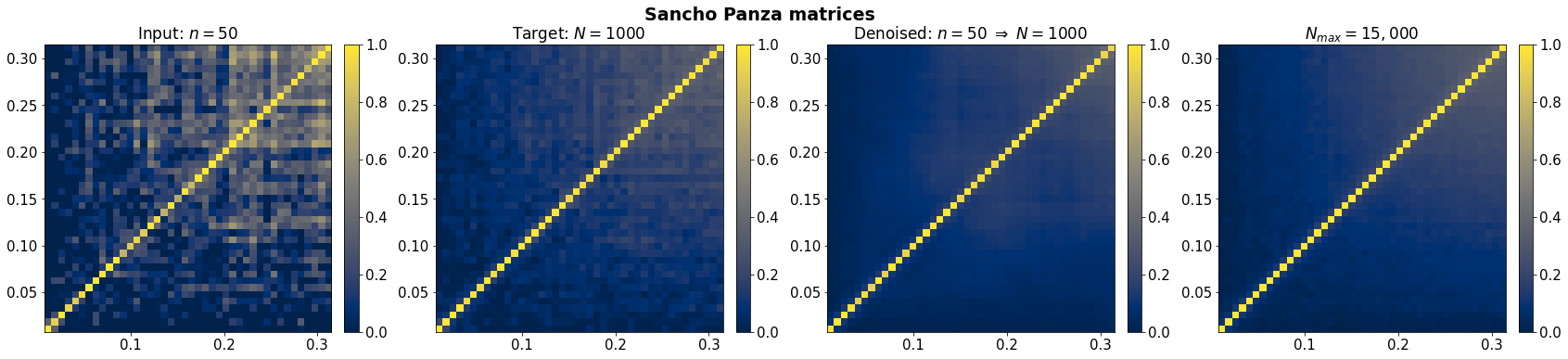}
  \caption{Comparison of the (normalized) cosmological covariance matrices for ExSHalos (first row) and Sancho Panza (last row). In the first column we have the input matrices with $n = 50$ spectra; the second column shows the target matrices, with $N = 1000$ spectra; the third column contains the respective denoised matrix, corresponding to the model for the combination of input and target samples $(n = 50, N = 1000)$; and
  the fourth column contains the {\em best} matrices, built using all the available spectra $N_{max}$. In each one of these figures, the axes corresponds to values of $k$, representing the $42$ Fourier bins of the power spectrum.
  }
  \label{fig:visual_matrices}
\end{figure}

In this section we present the main results of our work by comparing the input and denoised covariance matrices from the ExSHalos and Sancho Panza samples. 
The idea is, first, to measure by how much the ML denoiser
is able to remove the noise of the input covariance matrices in the ideal case, where we train, validate and test the method using only the ExSHalos sample.
The second step, and the key one, consists in applying the denoiser
trained with the help of ExSHalos to the input covariance matrices of the Sancho Panza samples, and comparing the denoised matrices to the best-case scenario, of a covariance matrix for a sample of many thousands of Sancho Panza simulations.
It is in this last phase that we are able to demonstrate the generalizing power of the suite, and to provide a proof-of-concept for the method.

We start with a visual presentation of the matrices in their normalized version.
We then compute the MSE between the cosmological covariance matrices obtained in the training stage (i.e., using the test subset of the ExSHalos matrices and all the trained models).
We also study the ranked eigenvalues of the matrices as a means to compare the loss of coherence due to noise, and how that is recovered by the denoiser.
In this section we also present an analytical approximation for the random process involved in the estimation of the sample covariance matrices in terms of the Wishart distribution, which also allows us to quantify the enhancement of the covariance matrices that comes about when applying the denoiser.
Finally, at the end of this section we present MCMC estimations of the cosmological parameters in the several different scenarios, showing by how much the denoiser
improves the accuracy and precision of the covariance matrices in terms of their end product -- the parameter constraints.
   
\subsection{Visualizing the matrices}
\label{sec:visual}

\begin{figure}[h!]
    \centering
    \includegraphics[scale=0.292]{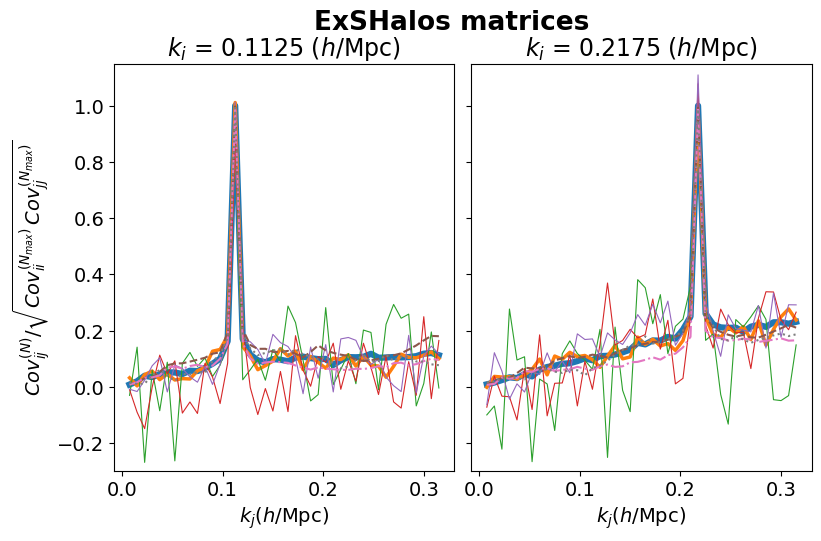}
    \hspace{-0.3cm}
    \includegraphics[scale=0.292]{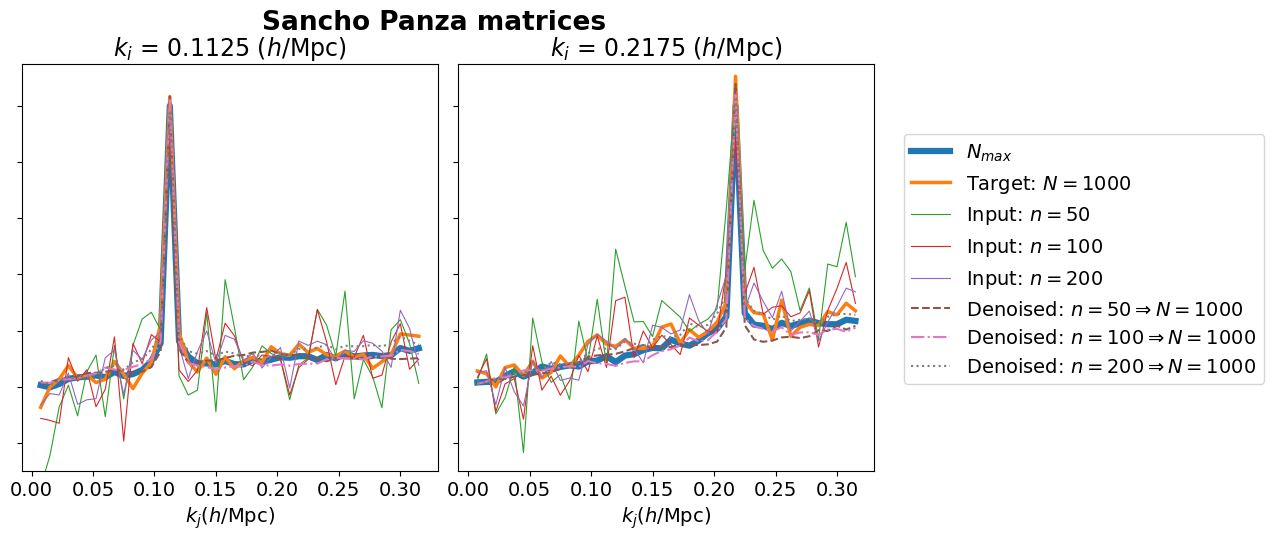}
    \caption{Slices of the normalized covariances (according to equation \ref{eq:normNbest}) of ExSHalos (left) and Sancho Panza (right), for different fixed values of $k_i$.
    The peaks correspond to points along the diagonal. 
    The plots show that the denoiser is able to remove the noise (seen in the input matrices), without introducing new features, so the denoised matrices match closely the targets and the {\em best} matrix (with a sample of $N_{max}$ spectra).}
    \label{fig:r}
\end{figure}

In figure \ref{fig:visual_matrices} we can already notice the power of the ML method from the visual representations of these matrices, by comparing the normalized matrices (according to equation  \ref{eq:normNbest}) against the target and the best-case scenario, where the covariance was computed with many thousands ($N_{max}$) of spectra.
The first row represents, from left to right, the input, target, denoised and best covariance matrices of ExSHalos, for the model with input and target matrices of sample sizes $(n = 50, N = 1000)$, respectively.
The second row represents the same, but for the Sancho Panza data set -- but we stress once again that the denoiser is exactly the same as the one in the first row, since it was trained only with the ExSHalos data set.
It can be seen from the figures that the denoised matrices appear almost identical as the best ones, which were obtained with all the spectra of the entire data set ($N_{max} = 30,000$ for ExSHalos and $N_{max} = 15,000$ for Sancho Panza).
This last feature can be explained by the choice of activation functions in the different layers, and because the ML models were trained using a huge number of different spectra, which appears to retain this information in the weights of the network.
Therefore, the denoised matrices are visually smooth and noiseless, especially when compared with the original, $n = 50$ covariance matrices.
At least from a purely visual standpoint, the power of the algorithm resides in the fact that the method is able to learn how to remove the noise using only ExSHalos matrices, and to apply this learning to the Sancho Panza ones.

A more accurate comparison of the matrices can be glimpsed from comparing slices (rows/columns) of the normalised covariance matrices, in order to show both the diagonal and off-diagonal elements \cite{Blot2019}. 
In figure \ref{fig:r} we show a few fixed $k_i$ slices of these matrices as a function of $k_j$, with the corresponding values for the  input, denoised, target and {\em best} normalised covariances. 
Roughly speaking, all the matrices follow the behaviour of the best-case scenario ($N_{max}$), but it is clear that the input matrices are severely affected with noise, especially in the off-diagonal elements and particularly for $n=50$ and $n=100$. 
However, after applying the ML denoiser those fluctuations basically disappear, and the off-diagonal elements match the behaviour of the target and {\em best} matrices in all cases, even for $n=50$. 
Moreover, from these plots it can also be seen that the off-diagonal structures in the ExSHalos matrices scale differently with $k$ compared with those of the Sancho Panza matrices, especially on small scales.
However, the denoiser (which was trained only with the ExSHalos matrices) is able to properly recover the off-diagonal behaviour of the Sancho Panza matrices as well, which is further evidence for the generalisation power of the ML method.

\subsection{The MSE between different matrices}
\label{sec:MSE}

\begin{figure}[h!]
  \centering
  \includegraphics[scale=0.75]{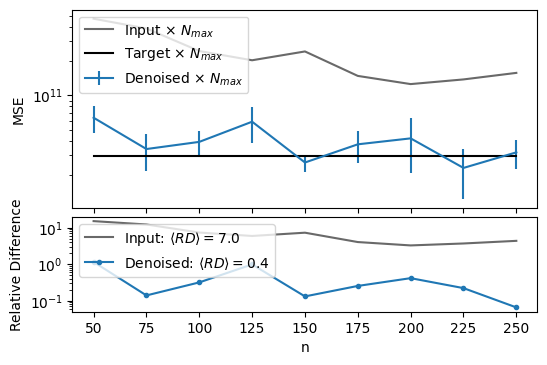}
  \caption{MSE for the cosmological covariance matrices, computed by comparing the best sample ($N_{max} = 30,000$) from ExSHalos and the original (input) matrices, in gray, and the denoised ones, in blue. 
  As the sample size of the input matrices ($n$) grow, the agreement between the matrices improve and the MSE becomes smaller. 
  The MSE between the best matrix and the target matrix (with $N = 1000$ spectra) is shown as the black line. 
  The error bars account for the values obtained for different seeds of the same model.
  The lower panel shows the relative difference according to equation \ref{eq:MSE}.
}
  \label{fig:MSE}
\end{figure}

Although the visual inspection of the previous subsection hints at the good performance of the method, we have monitored the improvements in the covariance matrices using the MSE metric, namely:
\begin{equation}
 MSE = \frac{1}{\mathcal{N}} \sum^{\mathcal{N}}_{l = 1} \frac{1}{n_k^2} \sum^{n_k}_{i, j = 1} (Cov^l_{i j} - Cov^{(N_{max})}_{i j})^2 \; , 
 \label{eq:MSE}
\end{equation}
where $Cov^l_{i j}$ are the input, target or denoised covariance matrices, 
$Cov^{(N_{max})}_{i j}$ is the best matrix (produced with the entire data set), $n_k$ the number of bins of $k$ in the data vector, and $\mathcal{N}$ is the total number of matrices used in the evaluation.
We computed the MSE only for the test subset of the ExSHalos matrices, and the results are shown in figure \ref{fig:MSE}.
The gray line corresponds to the MSE for the original (input) matrices,
the blue line corresponds to the denoised matrices, and the error bars account for the standard deviation for the results for each random seed.
As a comparison, the black line represents the MSE of the target matrices ($N=1000$), which is a lower bound for the MSE of the original matrices and provides a useful sanity check.
The lower panel in figure  \ref{fig:MSE} shows the residue:
\begin{equation}
    \label{Eq:RD}
  {\rm Relative \hspace{0.1cm} Difference} = \frac{abs\left( \text{MSE}_{\text{Y} \times 
  N_{max} } - \text{MSE}_{\text{Target} \times N_{max} } \right)}{\text{MSE}_{\text{Target} \times N_{max} }} ,
\end{equation}
where $Y$ stands for input or denoised. 
Naturally, the MSE decreases when the sample size ($n$) grows. 
The decrease in MSE that results from applying the denoiser is much larger: the mean values of the residue ($\langle RD \rangle$) show that the denoised matrices deviate by only $\sim 0.4$ from the best matrices, compared with approximately $7.0$ for the original input matrices -- a factor of more than 17 improvement.
Moreover, the MSE of the denoised matrices are weakly dependent on the original sample size, which may be indicative that even with an original sample of only $n=50$ spectra the denoiser
is already able to eliminate most of the noise of the covariance matrices, and increasing the sample size does not significantly improve upon that noise reduction. 
Notice that the black line (MSE of the target matrix) is reached in some cases by the denoised matrices, which is not completely unexpected since the training of the ML method includes more matrices than are included in the targets. 

\subsection{The eigenvalues of the matrices}
\label{sec:eigen}

\begin{figure}[h!]
  \centering
  \includegraphics[scale=0.55]{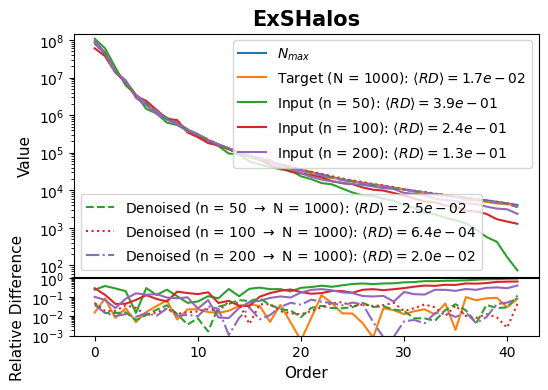}
  \includegraphics[scale=0.55]{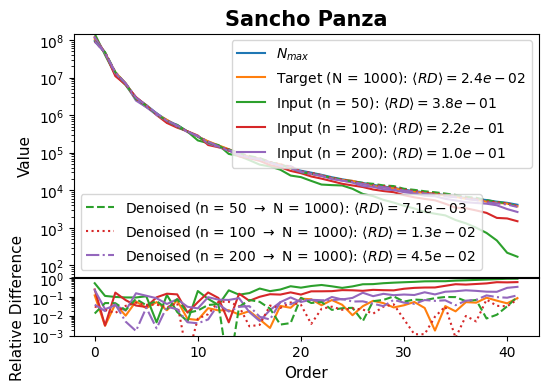}
  \caption{Ranked eigenvalues (main plot) and their relative difference (subplot, with the respective values for the matrix with $N_{max}$) for the ExSHalos (left) and Sancho Panza (right) covariance matrices.}
  \label{fig:eig_diag-ExSHalos}
\end{figure}

A complementary analysis to the one above can be made to compare the accuracy with which we reproduce the cosmological covariance matrices, which relies on the eigenvalues of those matrices.
Typically, the eigenvalues that effectively carry information obey some power law, and as we reach the lower eigenvalues the noise appears as an abrupt change in the scaling of the eigenvalues \cite{Vogeley1996, Ferreira2021}.
In figure \ref{fig:eig_diag-ExSHalos} we show the ranked eigenvalues for covariance matrices from ExSHalos (left) and Sancho Panza (right). 
We can see that the eigenvalues of the denoised matrices are much closer to the target and best matrices, while the original input matrices show clear signs of noise in the lower end of the spectrum of eigenvalues, which become more important as we decrease the sample size.
We quantify the difference in eigenvalues with respect to the best covariance matrix in terms of the same relative difference that was defined in equation \eqref{Eq:RD}.  
For ExSHalos, the improvements are a factor of more than $\sim 10$, while for Sancho Panza the improvements are between $\sim 50$ (for $n = 50$) and more than $\sim 10$ (for $n = 100$).
In a similar exercise, we have also compared the diagonal values of the non-normalized matrices -- see Appendix \ref{sec:appendix}.

\subsection{An analytical comparison for the ML black box}
\label{sec:wishart}

The encouraging results above indicate that the denoiser is effectively learning about the specific patterns of signal and noise, producing covariance matrices that are equivalent to those computed with samples of a much greater size than the sample size of the covariance that we plug into the
denoiser. In other words, applying the denoiser seems equivalent to increasing the sample size.

In order to check this conjecture we can compare the resulting (denoised) matrices with a model for the probability distribution function behind covariance matrices, and which describes how their fluctuations depend on the sizes of the data vector and the sample size.
There is, in fact, such an analytical description for covariance matrices in terms of the Wishart distribution \cite{Wishart1928, taylor2013}:
\begin{equation}
  p (\hat{M} | M, \nu, \eta) = \left( \frac{\nu^{\nu \eta /2} |M|^{- \nu/2} |\hat{M}|^{\gamma/2} }{ 2^{\nu \eta/2} \Gamma_{\eta} [\nu/2] } \right) \exp \left[ - \frac{\nu {\rm Tr} \left( \hat{M} M^{- 1} \right)}{2} \right] \; .
\end{equation}
Here, $M$ represents the statistical mean of the matrices, $|M|$ is its  determinant, and $\hat{M}$ is the random variable -- in this case, the sample covariance matrix. 
The parameters of this distribution are the size of the data vector
$\eta$ (the dimension of the matrices is $\eta \times \eta$), the number of degrees of freedom $\nu$ (in our case, the sample size).
In the formula above, $\gamma = \nu - \eta - 1$ and $\Gamma_{\eta} [\nu/2]$ is the multivariate Gamma function. 
Therefore, given an ``ideal'' covariance matrix $M$, this distribution allows us to generate random covariance matrices corresponding to different sample sizes. 
We should also point out that the Wishart distribution is unbiased, since $\langle \hat{M} \rangle = M$.

\begin{figure}[h!]
  \centering
  \includegraphics[scale=0.55]{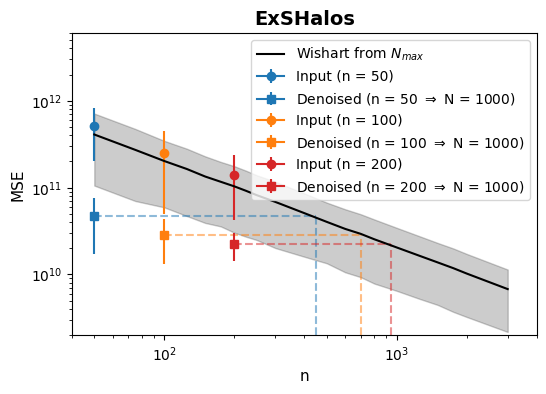}
  \includegraphics[scale=0.55]{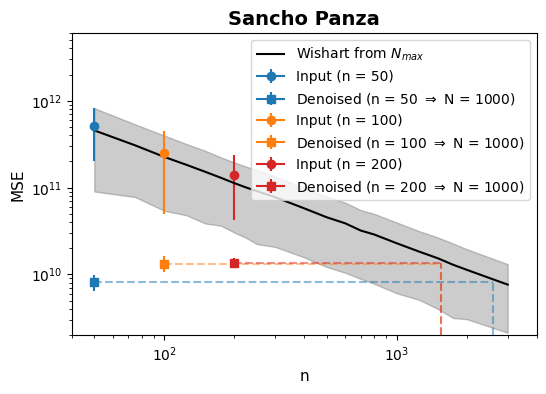}
  \caption{MSE comparison between the best ExSHalos (on the left) and Sancho Panza (on the right) matrices with: matrices estimated from the Wishart distribution, in black; the input, in circles (for $n \in [50, 100, 200]$); and denoised matrices, in squares (from $n \in [50, 100, 200] \Rightarrow N = 1000$). The colors corresponds to the number of matrices in the input and resulted denoised matrices. The gray region corresponds to $1 \sigma$ deviation for the mean values, in the case of the Wishart matrices. The dashed lines has the intention to guide the reader to see to which number of spectra $n$ the input matrices were taken to their Wishart comparison.}
  \label{fig:wishart}
\end{figure}

The Wishart distribution is useful in this context since it provides a model comparison
for the statistical fluctuations of covariance matrices as a function of sample size ($\nu \to n$). 
This means that we can infer what is the effective sample size of the denoised matrices by analyzing their MSEs, and comparing that with the result from the Wishart distribution where the ``ideal'' covariance matrix is computed using the maximum number of spectra available ($N_{max}$).
In figure \ref{fig:wishart} we plot the MSE (mean and variance) between the input and covariance matrices, compared with the covariance matrix obtained with all the spectra available ($N_{max} = 30,000$ for ExSHalos, and $N_{max} = 15,000$ for Sancho Panza).
The Wishart distribution (mean and variance) is denoted by the black line and gray region.
The input matrices are represented by the circles and their respective 1 $\sigma$ deviations, while the denoised matrices are represented by squares, both in the cases $n \in [50, 100, 200]$.
Since the mean MSE values for the denoised matrices are significantly lower compared with the input matrices, according to the Wishart distribution the denoiser is effectively taking matrices with samples of $n=50-200$ spectra, and transforming them into covariance matrices with much larger samples.

For the matrices from ExSHalos, the denoised matrices with $n \in [50, 100, 200]$ are closer to the prediction from the Wishart distribution for $n \in [450, 700, 940]$ spectra, respectively.
In the case of the Sancho Panza matrices, that effective sample size is even higher: $\sim 1550-2500$.
The denoised matrices also have a lower scatter, especially in the case of Sancho Panza, which means that the chance that a denoised sample covariance ends up producing a poorly estimated covariance matrix is lower than that for the equivalent effective sample size without denoising. 

It is also interesting that the Sancho Panza matrices benefit more from the denoiser.
This may be due to the fact that the matrices from Sancho Panza are already smoother than the ExSHalos matrices, resulting in better estimations to begin with. We have made tests changing the number of spectra in the normalization matrix of ExSHalos (to train the models), but still, the Sancho Panza results were better than for ExSHalos. Therefore, the normalization is not the cause behind this effect.

The interpretation presented in this section relies on a simple treatment of the covariance matrix, which is regarded as a quadratic combination of Gaussian random variables (in our case, the spectra).
A more thorough analysis can be made in terms of the likelihood, which takes into account a marginalization over the inverse Wishart distribution, leading to a modified multivariate $t-$distribution instead of a Gaussian distribution for the data \cite{Sellentin2016}. 
Similarly, using the Wishart distribution to propagate the uncertainty in the theoretical model, a Bayesian approach can be used to combine simulated and theoretical covariance matrices, also reducing the number of simulations required to reach some threshold of precision and accuracy \cite{Hall2019}.

\subsection{Recovering the cosmological parameters}
\label{sec:rec_pars}

\begin{figure}[h!]
  \centering
  {\bf \scriptsize ExSHalos}\\
  \hspace{0.2cm}
  \includegraphics[scale=0.4]{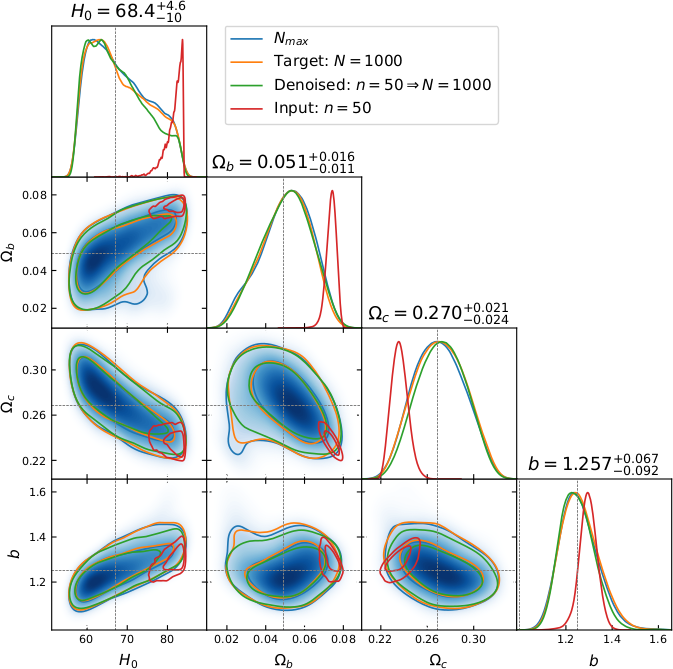}
  \hspace{0.2cm}
  \includegraphics[scale=0.4]{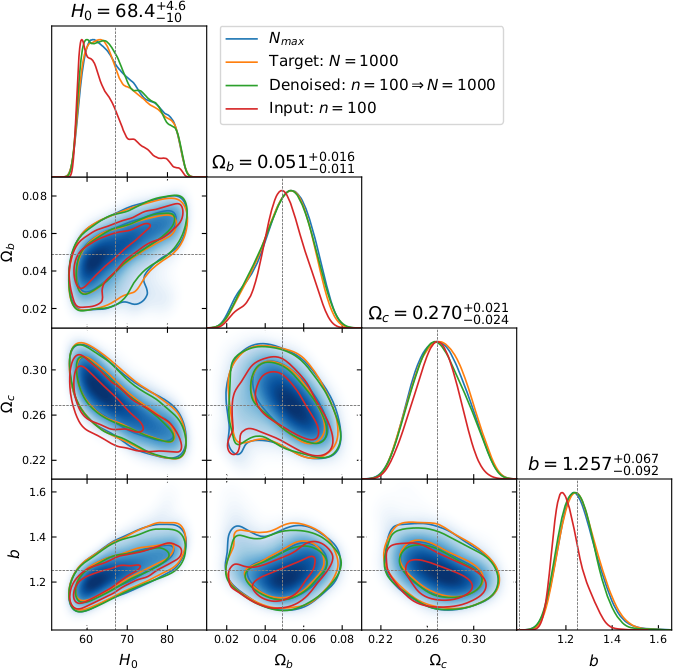}\\
  {\bf \scriptsize Sancho Panza}\\
  \hspace{0.2cm}
  \includegraphics[scale=0.4]{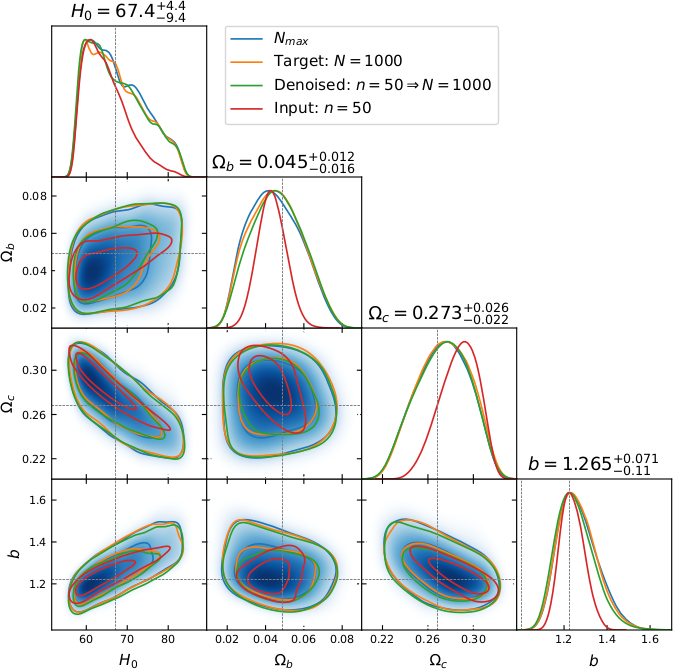}
  \hspace{0.2cm}
  \includegraphics[scale=0.4]{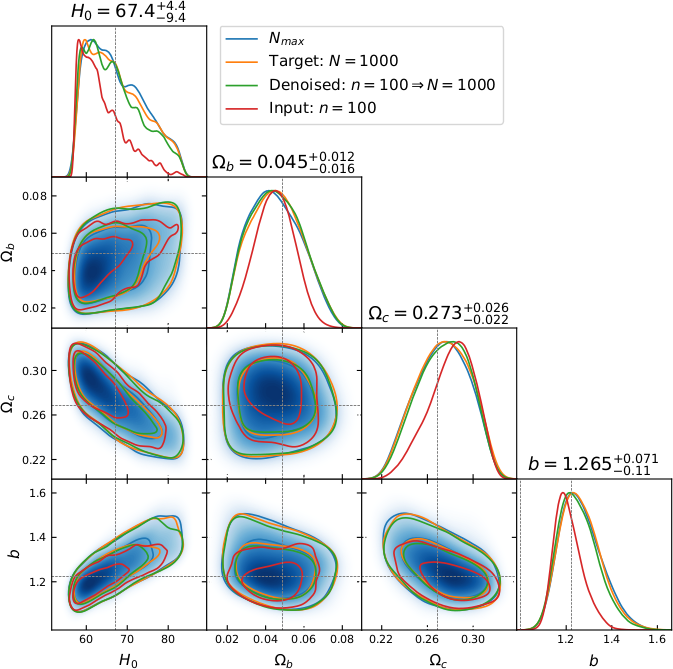}
  \caption{Cosmological parameter estimation comparison for different covariance matrices from ExSHalos (upper plots) and Sancho Panza (lower):
    matrices built using all the spectra available (with $N_{max} = 30,000$ spectra for ExSHalos, and $N_{max} = 15,000$ spectra for Sancho Panza); targets ($N = 1000$ spectra); input (using $n = [50, 100]$ spectra); and the denoised ones ($n \Rightarrow N$). 
    \label{fig:par_est}
    }
\end{figure}

Finally, it is important to analyze the ability of the denoised matrices to recover the fiducial
simulated parameters and compare these estimation with the parameters coming from the original matrices to validate our results.
The analysis is presented in figure \ref{fig:par_est}. 
We have explored the parameter space using the Markov Chain Monte Carlo (MCMC) approach, implemented with the help of the \href{https://emcee.readthedocs.io/en/stable/}{emcee} library \cite{emcee2013}, using as data points some random power spectrum vector, and using in each case the different cosmological covariance matrices. Our goal is to study the multivariate probability distribution function for the parameters: $H_0$, $\Omega_b$, $\Omega_c$ and the nuisance
parameter bias $b$, for the matrices built with the maximum number of spectra, target, denoised and input matrices. 
In all the analyses we have used 20 walkers and chains of 5000 length (except for the input matrices, for which we have used 6000).

Overall, in all models (for $n = 50$ or $n = 100$ and $N = 1000$) and both sets of matrices
(ExSHalos and Sancho Panza) all the parameters were well constrained using the denoised
matrices.
It is interesting to see that the inference considering the input matrices have a ``false''
precision (the volumes in the parameter space are lower, when compared to all the other matrices)
and is very inaccurate, because the mean values estimated is sightly (for $n = 100$) and highly 
(for $n = 50$, specially in the ExSHalos input matrix) shifted.
Moreover the input matrix with $n = 50$ and $n = 100$ spectra presents fluctuations on their contours, which are improved/removed in the denoised matrix estimation.

\begin{table}[h!]
 \caption{\label{tab:mean_comp} Bias in the estimated parameters, in units of the variance ($\mathbf{\Delta}$), measured with respect to the parameters determined using the covariance matrix built using all the spectra available ($N_{max}$).}
 \begin{center}
  \begin{tabular}{c|c|cc|cc}
   \cline{2-6}
   & \multirow{2}{*}{\textbf{Parameter}}& \multicolumn{2}{c|}{\textbf{ExSHalos}} & \multicolumn{2}{|c}{\textbf{Sancho Panza}} \\ 
   \cline{3-6}
   &  & \textbf{Input} & \textbf{Denoised} & \textbf{Input} & \textbf{Denoised} \\
   \cline{1-6}
   \multirow{4}{*}{n = 50} & $H_0$ & 1.8102 & 0.1019 & 0.3798 & 0.0406\\
    & $\Omega_b$ & 1.7542 & 0.0137 & 0.1066 & 0.1115\\
    & $\Omega_c$ & 1.5754 & 0.1037 & 0.6150 & 0.0090\\
    & $b$ & 0.4695 & 0.0889 & 0.2928 & 0.1442\\
   \cline{1-6}
   \multirow{4}{*}{n = 100} & $H_0$ & 0.4627 & 0.0443 & 0.4998 & 0.0953\\
    & $\Omega_b$ & 0.1177 & 0.0208 & 0.0020 & 0.0425\\
    & $\Omega_c$ & 0.1319 & 0.0097 & 0.3574 & 0.0659\\
    & $b$ & 0.5516 & 0.0917 & 0.6051 & 0.1254\\
   \cline{1-6}
   \multirow{4}{*}{n = 200} & $H_0$ & 0.1345 & 0.0075 & 0.0630 & 0.0124\\
    & $\Omega_b$ & 0.1510 & 0.0544 & 0.1245 & 0.0095\\
    & $\Omega_c$ & 0.1634 & 0.0627 & 0.0961 & 0.0191\\
    & $b$ & 0.4237 & 0.0318 & 0.0266 & 0.0350\\
   \cline{1-6}
   \end{tabular}
  \end{center}
\end{table}

Quantitatively, the improvements in the mean values becomes clear when we compare the bias in the expectation value of a parameter in units of its variance:
\begin{equation}
 \mathbf{\Delta} = \frac{\Delta \mu}{\sigma_{N_{max}}} = \frac{{\rm abs} \left( \mu_X -  \mu_{N_{max}}\right)}{\sigma_{N_{max}}} , \label{eq:DELTAO}
\end{equation}
where $\mu$ corresponds to the mean of a parameter, $X$ represents the input and denoised
matrices and $\sigma_{N_{max}}$ is the standard deviation of the parameter obtained using the {\em best} covariance matrix.
The results for ExSHalos and Sancho Panza matrices are presented in table \ref{tab:mean_comp}, in the scenarios where the input and denoised matrices are based on samples of $n = [50, 100, 200]$ spectra.

In the case of ExSHalos, the results of this quantity from the bad to the denoised matrices represent the improvements that is $\sim$ 128 times, in the case of $\Omega_b$, for models with $n = 50$; $\sim$ 13.6 times, in the case of $\Omega_c$, for the models with $n = 100$; and $\sim$ 17.9 times, in the case of $H_0$, for models with $n = 200$.
For matrices from Sancho Panza, the same comparison follows to $\sim 68$ times, for $\Omega_c$, for models with $n = 50$; $\sim$ 5.4 times, in the case of $\Omega_c$, for $n = 100$;
and  $\sim$ 13.1 times, in the case of $\Omega_b$, for $n = 200$.
The only parameters which are not improved are: $\Omega_b$, in the case of the model using $n = [50, 100]$; and $b$, for the model using $n = 200$, in the Sancho Panza analysis.
Notwithstanding, it is clear the improvement power of the proposed suite: using only about a hundred spectra for an input matrix, the denoiser can achieve results as if this matrix was made with thousands of spectra.
Moreover, the results for Sancho Panza completes the proof of the power of generalization of the proposed suite.

\section{Discussion and Conclusions}
\label{sec:conc}

The search for more efficient ways to compute the highly accurate cosmological covariance matrices that are now needed in cosmology has many different approaches
\cite{taylor2013, Meiksin1999, Tukey1958, Efron1980, Heavens2000, heavens2017, Philcox2021, Schneider2011, Scoccimarro2002, Chuang2015, PINOCCHIO2013, Kitaura2014, HALOGEN2015, Izard2016, Lognormal2017, ExSHalos2019, Chartier2021, Chartier2022}.
The main goal of this field is to provide precise matrices for parameter inference, that can be used in MCMC explorations of the likelihood, and result in unbiased estimates of the probability distribution function of the parameters.
The results of these analyses can lead to improved cosmological probes and/or experiments, for
different parameters; they can be used to break the tension of some parameters (as in the case of Hubble
parameter \cite{planck2018, Reid2019}); as well as in the search for the nature of dark energy.

This work presents an efficient approach for the estimation of cosmological covariance matrices.
The main idea behind our method is that, starting from matrices built with only hundreds of spectra, we are able to provide covariance matrices that are as good as if they were built with thousands of spectra.
We have implemented this method using CNNs and a denoising algorithm, that cleans the noise in the input matrices.
Visual inspection (see figures \ref{fig:visual_matrices} and  \ref{fig:r}) already shows that the noise was removed, without the introduction of any visible artifacts when compared with the best matrices.

The ML was trained in a data set of cosmological covariance matrices built with a mock generator of halo catalogues, called ExSHalos \cite{ExSHalos2019}. The reason behind this choice is the fact that this
code was already designed to be extremely fast, when compared to other mock generators or N-body simulations, and the ease of applying halo-finders in that context.
We were able to quickly generate many thousands of different maps, allowing us to build a very large data set of power spectra with which we could build our covariance matrices. 
In practice, the main limitation of our model comes from the computational cost associated with running the ExSHalos simulations. 
However, we should also note that the number of catalogues that were needed is rather small, since very good results can be obtained already for the model with $n = 50 \Rightarrow N = 1000$, which needs only 6000 ExSHalos maps. 
Moreover, as long as the ML models are trained, highly accurate covariance matrices from samples of N-body simulations can be obtained in a matter of seconds.

Although the ML suite was built using matrices coming from an approximated method, the main objective of this work was to test the generalization power of the suite when applying the models to matrices coming from N-body simulations such as Quijote \cite{QUIJOTE2020}. 
Hence, we are interested in taking the cleaning ML process learned with the covariance matrices from the approximated method, and applying it to matrices coming from the best simulations available. 
In this way, we expected to provide matrices that can even circumvent the well known errors (up to 10\%) in the parameter inference, that occur when using matrices that were derived only from mocks or other approximated methods \cite{Lippich2019, Blot2019, Colavincenzo2019}.

In order to check whether the resulting (denoised) covariance matrices were more similar to the best ones, we performed a series of tests. 
We started by computing the mean square error (MSE) of equation \eqref{eq:MSE}, and verified that the denoiser is able to reduce that indicator by a  factor of $\sim 10$ -- see figure \ref{fig:MSE}.
We also compared the ranked eigenvalues of the matrices, and showed that after denoising the input matrices we recover the main features of the {\em target} and even of the {\em best} matrices, which are built using the maximum number of spectra available ($N_{max}$).

An interesting point of debate is whether the predictions of a ML model can be compared with a simple
mathematical model. We have shown that the proposed method can be matched
in terms of an extrapolation according to the Wishart distribution \cite{taylor2013, Wishart1928}, by effectively augmenting the size of the sample that underlies the covariance matrix. 
Namely, in the case of the ExSHalos data set, the denoised covariance matrices built from an initial sample of $n=50-200$ spectra are equivalent to the ones computed using $\sim [450-940]$ spectra.
In the case of Sancho Panza the denoised matrices are similar to the ones computed with
$\sim [1550, 2500]$ spectra. 
The former result was expected, since the effect of the denoiser was to bring the matrices very close to the target matrices (with samples of $N=1000$). 
The latter result can be explained by the fact that the {\em best} matrix built with $N_{max} = 15,000$ Sancho Panza spectra has half as many spectra when compared ExSHalos (for which $N_{max} = 30,000$), and because those matrices are, indeed, smoother than the ones from ExSHalos. 

Ultimately, the strongest evidence for the power of the denoising technique is provided by the parameter estimation, in particular the plots in figure \ref{fig:par_est}, and the analysis of the bias as expressed by $\mathbf{\Delta}$ -- see equation \ref{eq:DELTAO} and table \ref{tab:mean_comp}. 
All the parameters ($H_0$, $\Omega_b$, $\Omega_c$ and $b$) were well
constrained in the case of the denoised matrices.
We see improvements (when comparing the input and denoised matrices) for all parameters in the case of ExSHalos, and for most parameters in the case of Sancho Panza.
In particular, the improvements achieved for $H_0$ using the sample with $n=50$ spectra were of a factor of $\sim 17.9$ in the case of ExSHalos, and of $\sim 9.4$ in the case of Sancho Panza.
Only when a parameter was already very well determined using the input covariance matrices ($\Delta \lesssim 0.1$), the denoised matrices did not lead to further improvements.
These results form the basis for establishing the predictive power of the proposed suite, as well as the generalization of the methods, which can be trained in covariance matrices from mocks and applied to covariance matrices from N-body simulations.

To summarize, in this work we have demonstrated the power of the use of image denoising techniques as a new approach to enhance sample cosmological covariance matrices. 
Our work already includes some modeling of real-life effects that are analogous to observational conditions (in terms of a non-trivial mask). 
Besides the architecture presented here, we have also tested different network configurations, different normalization methods to normalize the matrices, and even the residual encoder-decoder networks \cite{Mao2016}. 
In all these tests, the architecture presented in this work was the one which provided the best results.

The next steps regarding this project are: (i) to test our machinery in matrices from different cosmologies than the fiducial one, that was used in the training stage (to check if the method can be generalized also in those situations); (ii) to investigate if other ML image denoising methods can also be employed to improve the covariance matrices; and (iii) to apply the ML suite in more complex and realistic covariance matrices, e.g. in redshift space, with multiple tracers of the large-scale structure, and in higher-order statistics such as the bispectrum and trispectrum.

\acknowledgments

We would like to thank Elena Sellentin, Francisco Villaescusa-Navarro, Andrés Balaguera
Antolínez and Rodrigo Voivodic and Francisco Germano Maion for several useful comments. We
also thank the São Paulo Research Foundation (FAPESP),  the Brazilian National Council for
Scientific and Technological Development (CNPq), and the Coordination for the Improvement of
Higher Education Personnel (CAPES) for financial support. 
NSMS acknowledges financial support from FAPESP, grant
\href{https://bv.fapesp.br/pt/bolsas/187647/matrizes-de-covariancia-cosmologicas-e-metodos-de-machine-learning/}{2019/13108-0}.

\appendix
\section{The diagonal values of the matrices}
\label{sec:appendix}

\begin{figure}[h!]
  \centering
  \includegraphics[scale=0.55]{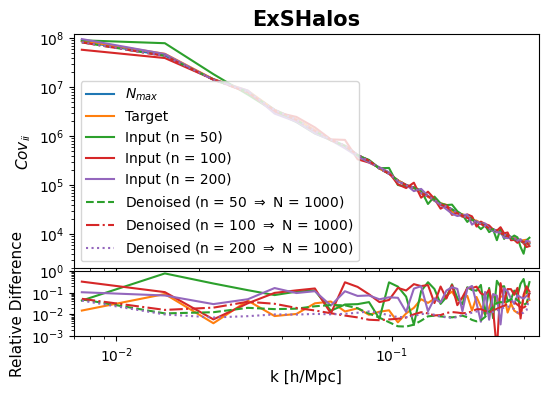}
  \includegraphics[scale=0.55]{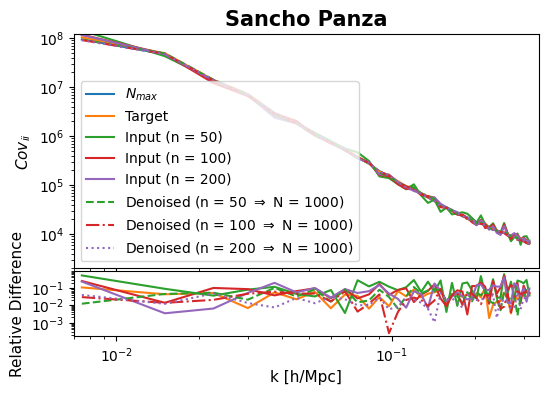}
  \caption{Diagonal values of the covariance matrices for the {\em best} case scenario ($N_{max}$ spectra), target ($N=1000$), input and denoised with $n = [50, 100, 200]$. The relative differences (lower subplot) are computed with respect to the {\em best} matrices.}
  \label{fig:diag}
\end{figure}

As an additional check to ensure that the denoised matrices match the target or the {\em best} matrices, we have also looked at the values of the diagonals of those matrices.
In figure \ref{fig:diag} we show that comparison for the {\em best} matrices (computed with a sample of $N_{max}$ spectra), the target ($N=1000$), as well as the input and denoised matrices in the cases of samples $n=[50,100,200]$.
In both cases (ExSHalos and Sancho Panza) the diagonals of the denoised matrices are a much better match to the diagonal of the {\em best} covariance matrix, with relative differences lower than $0.1$.
This is in contrast with the input (noisy) matrices, which even for $n=200$ still show deviations greater than 10\% compared with the {\em best} matrix.
This result, combined with the comparison of the eigenvalues of section \ref{sec:eigen}, shows that the denoiser is able to recover the key features of the matrices, leading to denoised covariances which are at least as good as the target, and often come very close to the best case scenario.


\end{document}